\documentclass[aps,prl,reprint,superscriptaddress]{revtex4-1}
\usepackage{siunitx}
\usepackage{mhchem}
\usepackage{bm}
\usepackage{graphicx}
\usepackage[colorlinks]{hyperref}
\usepackage{amsmath}
\usepackage{amssymb}
\usepackage{bbm}
\usepackage{times}
\usepackage[normalem]{ulem}

\hypersetup{citecolor=blue}

\newcommand{\figurewidthwide}{.90\textwidth}
\newcommand{\figurewidthhalf}{.48\textwidth}

\usepackage{xr}
\begin{document}

\title{Active learning of effective Hamiltonian for super-large-scale atomic structures}

\author{Xingyue Ma}
\affiliation{National Laboratory of Solid State Microstructures and Collaborative Innovation Center of Advanced Microstructures, Nanjing University, Nanjing 210093, China}
\affiliation{Jiangsu Key Laboratory of Artificial Functional Materials, Department of Materials Science and Engineering, Nanjing University, Nanjing 210093, China}

\author{Hongying Chen}
\affiliation{National Laboratory of Solid State Microstructures and Collaborative Innovation Center of Advanced Microstructures, Nanjing University, Nanjing 210093, China}
\affiliation{Jiangsu Key Laboratory of Artificial Functional Materials, Department of Materials Science and Engineering, Nanjing University, Nanjing 210093, China}

\author{Ri He}
\affiliation{Key Laboratory of Magnetic Materials Devices \& Zhejiang Province Key Laboratory of Magnetic Materials and Application Technology, Ningbo Institute of Materials Technology and Engineering, Chinese Academy of Sciences, Ningbo 315201, China}

\author{Zhanbo Yu}
\affiliation{National Laboratory of Solid State Microstructures and Collaborative Innovation Center of Advanced Microstructures, Nanjing University, Nanjing 210093, China}
\affiliation{Jiangsu Key Laboratory of Artificial Functional Materials, Department of Materials Science and Engineering, Nanjing University, Nanjing 210093, China}

\author{Sergei Prokhorenko}
\affiliation{Smart Functional Materials Center, Physics Department and Institute for Nanoscience and Engineering, University of Arkansas, Fayetteville, Arkansas 72701, USA}

\author{Zheng Wen}
\affiliation{College of Electronics and Information, Qingdao University, Qingdao, 266071, China}

\author{Zhicheng Zhong}
\affiliation{Department of Physics, University of Science and Technology of China, Hefei 230026, China}
\affiliation{Suzhou Institute for Advanced Research, University of Science and Technology of China, Suzhou 215123, China}

\author{Jorge Íñiguez-González}
\affiliation{Materials Research and Technology Department, Luxembourg Institute of Science and Technology (LIST), Avenue des Hauts-Fourneaux 5, L-4362 Esch/Alzette, Luxembourg}
\affiliation{Department of Physics and Materials Science, University of Luxembourg, 41 Rue du Brill, L-4422 Belvaux, Luxembourg}

\author{L. Bellaiche}
\affiliation{Smart Functional Materials Center, Physics Department and Institute for Nanoscience and Engineering, University of Arkansas, Fayetteville, Arkansas 72701, USA}

\author{Di Wu}
\affiliation{National Laboratory of Solid State Microstructures and Collaborative Innovation Center of Advanced Microstructures, Nanjing University, Nanjing 210093, China}
\affiliation{Jiangsu Key Laboratory of Artificial Functional Materials, Department of Materials Science and Engineering, Nanjing University, Nanjing 210093, China}

\author{Yurong Yang}
\affiliation{National Laboratory of Solid State Microstructures and Collaborative Innovation Center of Advanced Microstructures, Nanjing University, Nanjing 210093, China}
\affiliation{Jiangsu Key Laboratory of Artificial Functional Materials, Department of Materials Science and Engineering, Nanjing University, Nanjing 210093, China}

\date{\today}

\begin{abstract}

  The first-principles-based effective Hamiltonian scheme provides one of the most accurate modeling technique for large-scale structures, especially for ferroelectrics. However, the parameterization of the effective Hamiltonian is complicated and can be difficult for some complex systems such as high-entropy perovskites.
  Here, we propose a general form of effective Hamiltonian and develop an active machine learning approach to parameterize the effective Hamiltonian based on Bayesian linear regression. The parameterization is employed in molecular dynamics simulations with the prediction of energy, forces, stress and their uncertainties at each step, which decides whether first-principles calculations are executed to retrain the parameters.
  Structures of \ce{BaTiO3}, \ce{Pb(Zr_{0.75}Ti_{0.25})O3} and \ce{(Pb,Sr)TiO3} system are taken as examples to show the accuracy of this approach, as compared with conventional parametrization method and experiments.
  This machine learning approach provides a universal and automatic way to compute the effective Hamiltonian parameters for any considered complex systems with super-large-scale (more than $10^7$ atoms) atomic structures.

\end{abstract}

\maketitle

First-principles (FP) methods based on density functional theory (DFT) has become indispensable to scientific research in physics, chemistry, materials and other fields \cite{DFT_martin_electronic_2004}. However, studying the structure and properties of large scale structures, such as thermally-driven phase transitions or multidomain states in ferroic materials, remains a great challenge due to the large computational cost using \emph{ab initio} molecular dynamics.
The recent development of first-principles-based machine learning force fields (MLFFs) for the molecular dynamics make it possible to study the large scale structure with good accuracy similar to first-principles \cite{VaspMLFF_PhysRevB.100.014105,maxvol_PODRYABINKIN2017171,FLARE_vandermause_fly_2020,MLFF_Review_jinnouchi_-fly_2020,MLFF_Review_friederich_machine-learned_2021}.
Another method that can handle large scale structure is the first-principles-based effective Hamiltonian,  which is also physically interpretable and faster than MLFF-based molecular dynamics.
The first-principles-based effective Hamiltonian approach has been proposed to describe the couplings between local order parameters (both long-ranged and short-ranged), in which the coupling parameters for the  effective Hamiltonian  are computed by first principles and have direct physical meanings \cite{BTO_Zhong1995,SPLD_review_ghosez_modeling_2022}. 
Such method has successfully reproduced or predicted the structure phase transitions \cite{BTO_PhysRevLett.73.1861,Topo_nahas_discovery_2015,KNO_chen_deterministic_2022,CPI_Chen2020}  and various properties of many compounds \cite{Dielec_NC_wang_subterahertz_2016,Piezo_NC_nahas_microscopic_2017,PST_PhysRevB.105.054104,KNO_chen_deterministic_2022,Nature_Geofrus_choudhury_geometric_2011}, such as piezoelectric effect, electrocaloric effect, dielectric response and optical response and so on. Moreover, interesting complex polar vortices \cite{HeffVort_Naumov2004}, ferroelectric labyrinthine domains \cite{Labr_Nahas2020}, polar skyrmion \cite{skyr_Han2022} and merons \cite{skyr_PhysRevLett.120.177601} were also recently found by effective Hamiltonian methods in complex perovskite systems.

For the effective Hamiltonian, the parameters of order-parameter-couplings are obtained by fitting FP calculations for many structures with special structural distortions \cite{1a,BTO_Zhong1995}. These fitting procedures may be tricky and complex, and some approximations (such as virtual crystal approximation) \cite{PZT_PhysRevLett.84.5427,BST_PhysRevB.73.144105,BST_PhysRevB.99.064111} may need to be included, leading to uncertainties and even errors for some complex interactions and structures. Additional manual adjusting the values of some parameters may be necessary to reproduce experimental results \cite{KNO_chen_deterministic_2022,CPI_Chen2020}.  Therefore, to avoid the complication and approximation in the parameterization of the effective Hamiltonian, a new scheme of parameterization in a reliable, precise, convenient, and automatic way is highly demanded.

In this article, a general effective Hamiltonian is proposed and on-the-fly active learning method is applied to the parameterization of this general effective Hamiltonian. Perovskite structures [\ce{BaTiO3}, \ce{Pb(Zr,Ti)O3}, and \ce{(Pb,Sr)TiO3}] are taken as examples, where the active learning method provides simulations results that agree very well with other first-principles-based calculations and experiments.
Such reliable and highly-automatic way to construct the effective Hamiltonian parameters makes it possible to mimic the superlarge scale and complex atomic structures. 



\section{Results}

\begin{figure*}
  \includegraphics[width=\figurewidthwide]{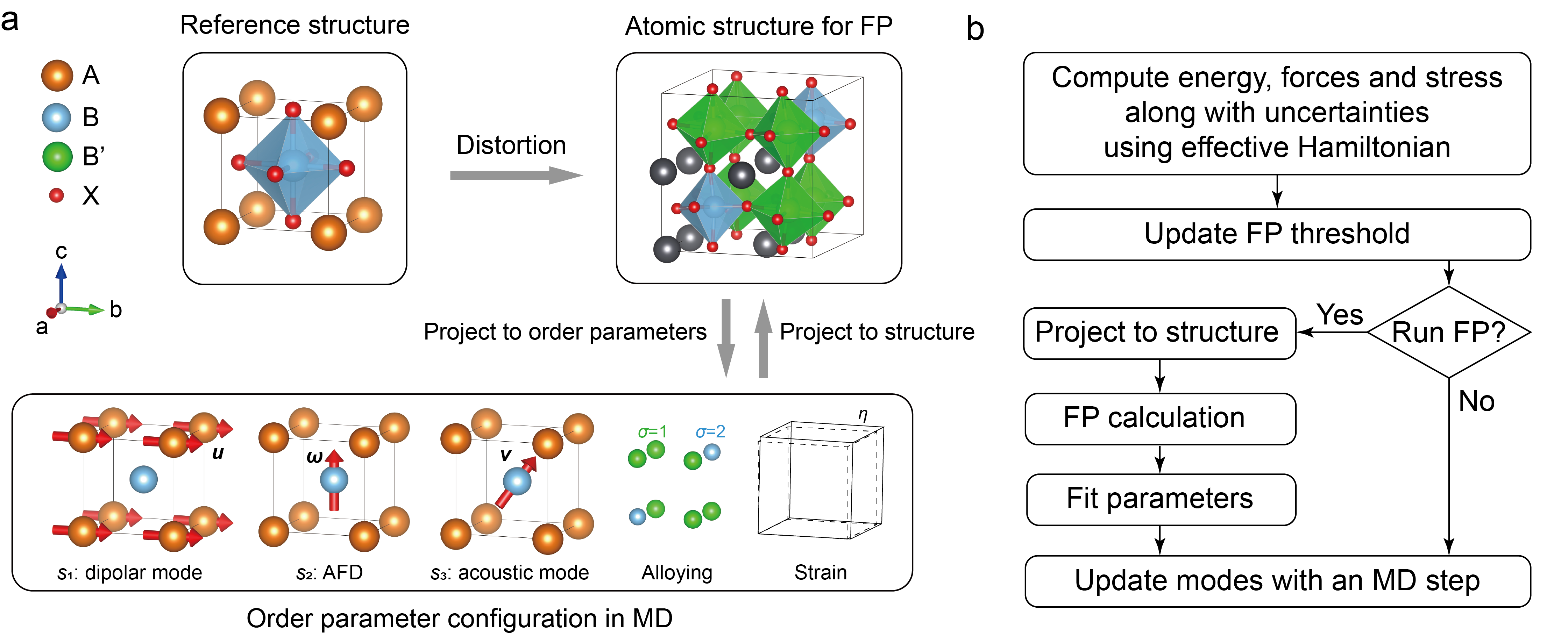}
  \caption{\label{fig:schematic} Schematics of on-the-fly learning of effective Hamiltonian for perovskite structure. (a) The projection between the atomic structure used in the FP calculations and the order-parameters configuration used in effective Hamiltonian MD simulations. (b) The workflow of on-the-fly learning of effective Hamiltonian. }
\end{figure*}

\subsection{Effective Hamiltonian}

The effective Hamiltonian describes the couplings between order parameters, and it is developed based on the Taylor expansion of small distortions around the reference structure.
Various order parameters are considered (see Fig. \ref{fig:schematic}a for an example), which is further explained in ``Methods''. Briefly, the degrees of freedom of the effective Hamiltonian are:
(1) local modes attributed to each unit cell $i$, to be denoted as $\lbrace \bm s_1\rbrace, \lbrace \bm s_2\rbrace, \cdots$, representing atomic displacements with respect to the reference structure, usually associated with different phonon modes;
(2) the homogeneous strain tensor $\eta$ \cite{BTO_Zhong1995}; and
(3) the variable $\lbrace\sigma\rbrace$ representing the atom occupation in unit cell $i$  [for example, in \ce{Pb(Zr,Ti)O3},  $\sigma_i=1$ (respectively, $\sigma_i=2$) represents a Ti (respectively, Zr) atom sitting in unit cell $i$]  \cite{PZT_PhysRevLett.84.5427}.

The potential energy $E_{\text{pot}}$ contains four main parts: (i) $E_{\text{single}}$, which contains the self energies of each mode, involving only one site in each term; (ii) $E_{\text{strain}}$, which contains all the energy terms directly related to the strain tensor $\eta$; (iii) $E_{\text{inter}}$, which contains several terms 
describing the two-body interactions between different local modes or the same local modes at different sites; and (iv) $E_{\text{spring}}$, which describes the effect of atomic configuration of different elements (known as ``alloying effect'' \cite{PZT_PhysRevLett.84.5427}), which consists of several ``spring'' terms (using the terminology of Ref. \cite{BZT_spring_PhysRevB.106.224109}). 

In principle, the formalism of effective Hamiltonian could be applied to any structures where a reference structure with high symmetry can be defined. Here, we mainly focus on the perovskites in formula \ce{ABX3}, in which A- or B- sites can be occupied by multiple elements. Practically, the local dipolar mode vector $\lbrace\bm u\rbrace$, antiferrodistortive (AFD) pseudovector $\lbrace\bm\omega\rbrace$ and anharmonic strain vector (acoustic mode) $\lbrace\bm v\rbrace$ are considered as the modes $\lbrace \bm s\rbrace$ (see Fig. \ref{fig:schematic}a, and in total nine degrees of freedom describe the state of each unit cell).
More details of the effective Hamiltonian for perovskites are described in ``Methods''.

\subsection{Formalism of the parametrization}
Instead of doing many FP energy calculations on \emph{special} structures with distortions to compute the coefficients of order-parameters coupling in effective Hamiltonian as in previous reports \cite{1a,BTO_Zhong1995}, our present approach is to use the on-the-fly active learning approach to  automatically compute the parameters for  effective Hamiltonian. The parameters related to the long-range dipolar interaction $E_{\text{long}}$ [Eq. \eqref{eq:long_range} in ``Methods''] [i.e. the lattice constant $a_0$, the dipolar mode Born effective charge $Z^*$ and optical dielectric constant $\epsilon_\infty$ (using the notations of Ref. \cite{BTO_Zhong1995})] are first determined directly from  first principles calculations. Then, all the remaining parameters are determined through on-the-fly machine learning process.
As indicated in Methods, the effective Hamiltonian can be written in the following form
\begin{equation}
  E_{\text{pot}} = E_{\text{long}} + \sum_{\lambda=1}^M w_\lambda
  t_\lambda (\lbrace \bm u\rbrace,\lbrace \bm v\rbrace,\lbrace \bm \omega\rbrace,\lbrace \sigma\rbrace, \eta),
  \label{eq:linear}
\end{equation}
where $E_{\text{long}}$ is fixed during the fitting process, $M$ is the number of parameters to be fitted, $w_\lambda$ is the parameter to be fitted, and $t_\lambda$ is the energy term associated with the parameter $w_\lambda$, which is called symmetry-adapted term (SAT), using the terminology of Ref. \cite{SPLD_PhysRevB.95.094115}. In other words, except long-range dipolar interactions, the energy of the system is \emph{linearly} dependent on the \emph{parameters}. Moreover, the force (respectively, stress) have similar forms to the energy, which is obtained by taking derivative over local mode (respectively, strain) on $E_{\text{long}}$ and $t_\lambda$.
Such linearity is similar to the second-principle lattice dynamics \cite{SPLD_PhysRevB.95.094115} and MLFF with Gaussian approximation potential \cite{GAP_PhysRevLett.104.136403}, allowing the application of similar regression algorithms. Here, we use the Bayesian linear regression algorithm similar to that previously used for  MLFF \cite{VaspMLFF_PhysRevB.100.014105}, with several key modifications for the effective Hamiltonian context.

Given the linearity above, the linear parts of energy, force and stress for each structure $a$ calculated from the effective Hamiltonian could be written in the following matrix form
\begin{equation}
  \tilde{\mathbf{y}}_a\equiv \mathbf y_a - \mathbf y_{a}^{\text{long}} = \bm {\phi}_a \mathbf w,
\end{equation}
where $\mathbf y$ is a vector containing the energy per unit cell with respect to the reference structure, the forces acting on the modes and the stress tensor  (in total $m_a=1+9N_a+6$ elements, where $N_a$ is the number of unit cells in the structure $a$, and we consider the above mentioned nine local modes in our models), $\mathbf{y}^{\text{long}}$ is a vector in similar layout associated with the $E_{\text{long}}$ term, and $\mathbf w$ is a vector that consists of all the parameters $w_\lambda, \lambda=1,\cdots,M$; and $\bm \phi_a$ is an $m_a \times M$ matrix consisting of the SATs and their derivatives with respect to modes and strain.
Note that in the effective Hamiltonian formalism, the potential energy of the reference structure is zero by definition. Thus, the energy obtained from the FP calculations in $\mathbf y$ should be subtracted by the energy of the reference structure to be consistent with the effective Hamiltonian.

In the parametrization process, a set of structures are selected as the training set (see ``On-the-fly learning''), and the structures are indexed by $a=1,\cdots,N_T$. First-principles calculations are performed on these structures to get the energy per unit cell, forces acting on the atoms, and the stress tensor. The forces acting on the modes are then obtained by applying Eq. \eqref{eq:force_map} (more details are described in ``Mode and basis'' in ``Methods''). The $\tilde{\mathbf y}_a$ vector of all the structures in the training set then constitutes the vector $\mathbf Y$ containing $\sum_a m_a$ elements. On the other hand, the $\bm \phi_a$ matrices of the structures in the training set constitute the $\bm \Phi$ matrix. In this form, the parametrization problem is to adjust $\mathbf w$ to fit $\bm \Phi \mathbf w$ against $\mathbf Y$.
To balance the energy, force and stress values with different dimensions properly, they are typically divided by their standard deviation in the training set to get dimensionless values.
Furthermore, an optional weight could be assigned to each of the types to adjust the preference between different fitting targets. Practically, this is achieved by left multiplying a diagonal matrix $\mathbf H$ made up by $h_i/\sigma_i$ to $\bm \Phi$ and $\mathbf Y$,
where $h_i$ and $\sigma_i$ are the weight and standard deviation of the specified type of values (energy, force of different modes and stress), respectively.

Once the training set is supplied, the parameters as well as their uncertainties could be predicted with the Bayesian linear regression (see ``Methods''). Moreover, for a new structure that is not contained in the training set (i.e. \emph{no} FP calculation is performed on such structure), the energy, forces and stress as well as their uncertainties could be predicted by the regression model. Such uncertainties are used to determine whether FP calculation is necessary for such structure.

\begin{figure*}
  \includegraphics[width=\figurewidthwide]{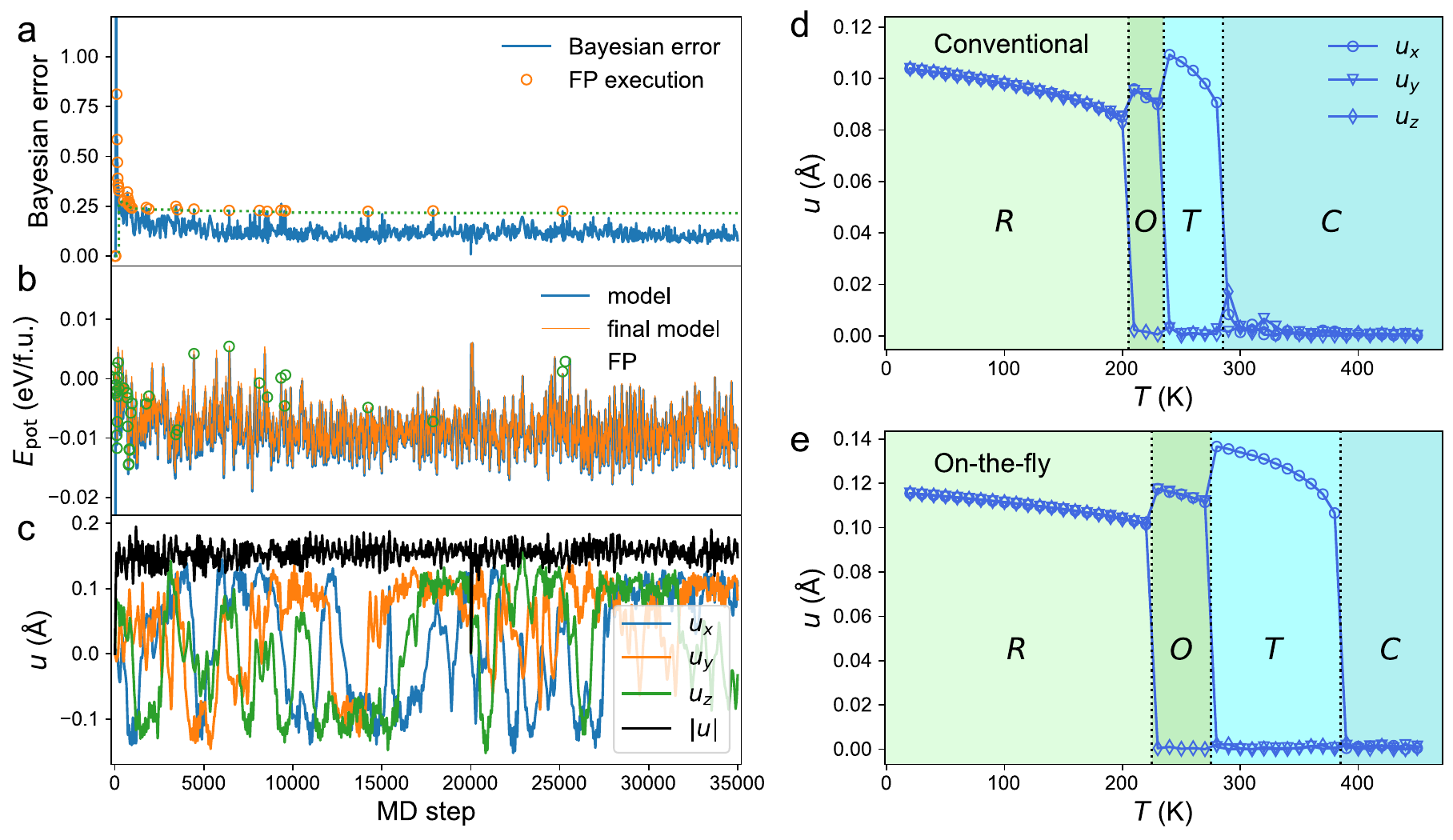}
  \caption{\label{fig:BTO_OTF}
  On-the-fly machine learning of parametrization for \ce{BaTiO3}. (a, b, c) (a) Bayesian error, (b) potential energy per formal unit (f.u.), and (c) local dipolar mode $\bm u$ in the learning/fitting process as functions of MD steps. The dash line in panel (a)  denotes the threshold to perform FP calculations. The blue and orange lines in panel (b) represent the energy computed by effective Hamiltonian model during the fitting process and with the the final parameter after the learning process, respectively.
  Phase diagram by effective Hamiltonian simulations with the parameters from (d) FP calculations in Ref. \cite{BTO_Zhong1995} and (e) on-the-fly learning. Absolute values of local mode $\bm u$ of \ce{BaTiO3} as functions of temperature are shown in panel (d) and (e). Here, R, O, T, and C denote the rhombohedral, orthogonal, tetrahedral and cubic phases, respectively.
  }
\end{figure*}

\subsection{On-the-fly learning}

In our approach, the parameters of effective Hamiltonian are fit in a scheme similar to that generating on-the-fly machine learning force field (MLFF) \cite{VaspMLFF_PhysRevB.100.014105} with some modifications for the effective Hamiltonian scheme. The parameters are fitted during effective Hamiltonian MD simulations 
on relatively small cells. The effective Hamiltonian MD simulations are performed by solving the equations of motions of each degree of freedom %
\begin{equation}
  \begin{split}
    \frac{\partial p_i}{\partial t} &= -\frac{\partial E_{\text{pot}}}{\partial s_i}, \\
    \frac{\partial s_i}{\partial t} &= \frac{p_i}{m_i},
  \end{split}
\end{equation}
where $p_i$ is the momentum associated with the degree of freedom $s_i$; $m_i$ is the effective mass of the degree of freedom $s_i$, which is obtained as in Ref. \cite{MD_nishimatsu_fast_2008}; and $t$ is the time.
As shown in Fig. \ref{fig:schematic}b, in each MD step, the energy, forces and stress tensor on the structure as well as their uncertainties are predicted by the effective Hamiltonian with the current parameters and collected data using the Bayesian linear regression.
If the uncertainty (Bayesian error) of the energy, forces or stress tensor is large, FP calculation is executed, the corresponding results are stored into the training set, and the parameters are refitted using the updated training set; otherwise, the FP calculation is skipped. Then, the structure is updated by executing one MD step with the forces and stress from the FP calculation (if available) or those from the effective Hamiltonian.

During the fitting process, the Bayesian errors of the energy, forces and stress predicted by the effective Hamiltonian are calculated by Eq. \eqref{eq:uncert_label} (see more details in Methods) and compared to the threshold to determine whether FP calculation is necessary.
At the beginning, the threshold is typically initialized with zero. Before the setting up of nonzero threshold, the FP calculations take place in fixed interval of several steps (say, 10 or 20 MD steps). The threshold is then updated dynamically during the fitting process using the flow similar to that in Ref. \cite{VaspMLFF_PhysRevB.100.014105}, with the exception that the spilling factor is not used in this work.
Note that different from Ref. \cite{VaspMLFF_PhysRevB.100.014105}, the parameters are typically fitted \emph{immediately} as soon as when new FP calculation is finished, instead of fitted after several FP results are obtained. This difference stems from the observation that the parameter fitting for the effective Hamiltonian is typically very fast compared to the FP calculations. Such immediate fitting is helpful for reducing the required number of FP calculations  and improve the fitting efficiency.

\subsection{Applications}

\begin{figure*}
  \includegraphics[width=\figurewidthwide]{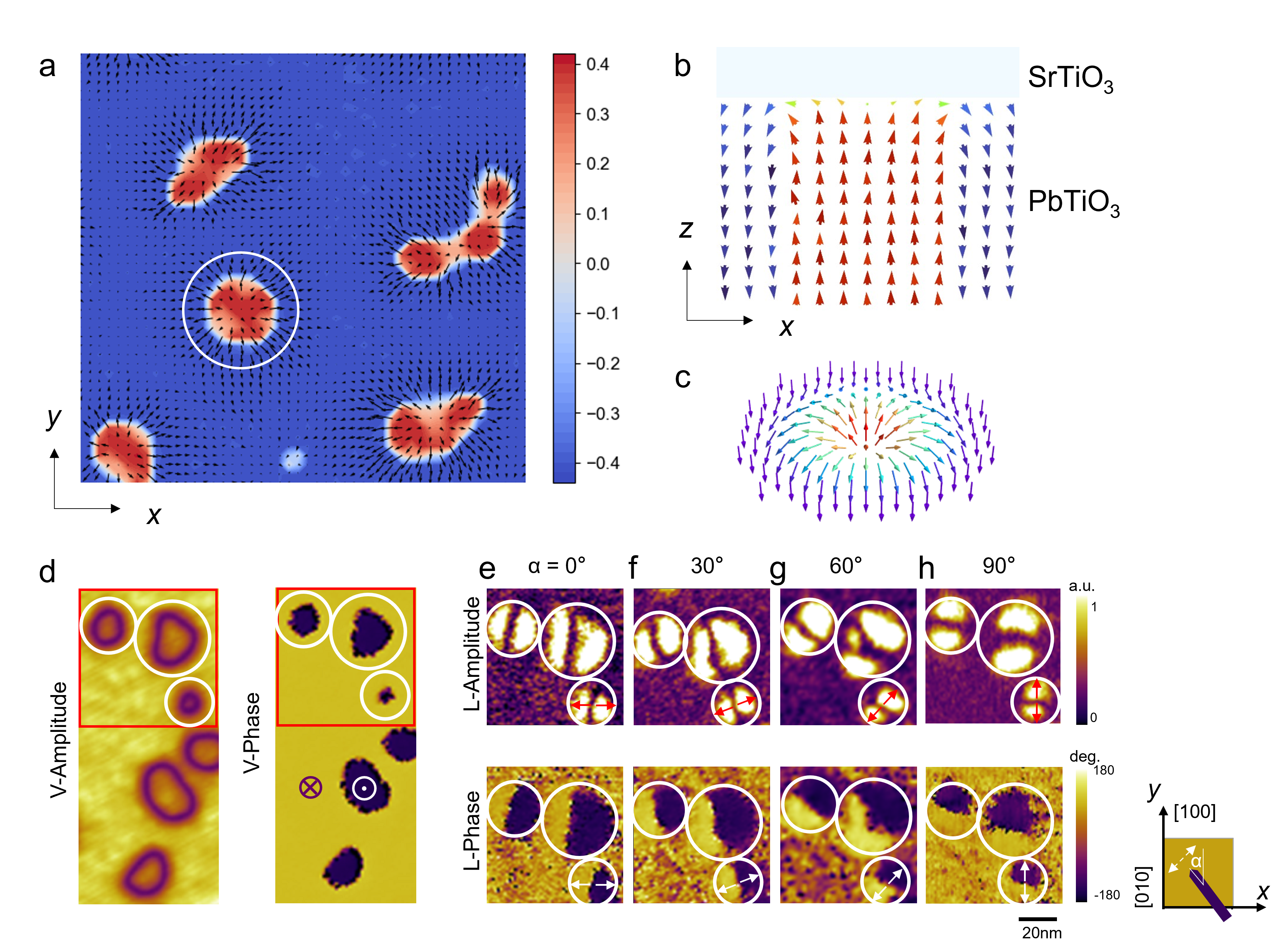}
  \caption{\label{fig:skyr} Polar distribution of \ce{SrTiO3}/\ce{PbTiO3} bilayer.
  (a) Dipole configuration of \ce{SrTiO3}/\ce{PbTiO3} averaged over the top 10 planes of \ce{PbTiO3} layer obtained from HMC simulation with the effective Hamiltonian. The arrows denote the in-plane component of the local dipolar mode, and the colors denote the out-of-plane components of dipolar modes.  
  (b) The dipole configuration of \ce{SrTiO3}/\ce{PbTiO3} in a (010) plane around the circled skyrmion in panel (a). The colors of the arrows denote the out-of-plane component of local dipolar mode.
  (c) The schematic of the skyrmion at the top of the \ce{PbTiO3} layer, where the color of the arrows denote the out-of-plane component of local dipolar mode.
  (d-h) The PFM images characterizing the skyrmion in \ce{SrTiO3}/\ce{PbTiO3}. 
  (d) Vertical PFM amplitude and phase. 
  (e-h) Lateral PFM amplitudes and phases corresponding to the range marked with red rectangles in panel (c), with tip orientation angles of \SI{0}{\degree} (e), \SI{30}{\degree} (f), \SI{60}{\degree} (g) and \SI{90}{\degree} (h). The white circles in panels (d-h) mark one of the skyrmion. The arrows in the circles in panels (e-h) mark the in-plane component of the polarization.
  }
\end{figure*}

\paragraph*{Simple perovskite \ce{BaTiO3}.}
The on-the-fly learning effective Hamiltonian is first applied to simple perovskite \ce{BaTiO3}, which is one of the most studied ferroelectric perovskites.
Figures \ref{fig:BTO_OTF} a, b and c show the parameter evolutions during on-the-fly machine learning.
The simulation is performed on $2\times 2\times 2$ supercell (40 atoms) at the temperature of \SI{50}{K}.
The Bayesian error during the fitting process is displayed in Fig.\ref{fig:BTO_OTF}a. At the beginning (about 500 steps), the Bayesian error is quite large, and FP calculations are called frequently. As the fitting progresses, more FP data is collected and the parameters are updated, leading to the rapid decline of Bayesian error. The threshold is also adjusted dynamically in this process. After about 1000 MD steps, the threshold is nearly unchanged and the FP calculations are only rarely required. Figure \ref{fig:BTO_OTF}b shows the potential energy predicted by the effective Hamiltonian and that computed from FP calculations in the simulation, showing they are close to each other at each step.
Figure \ref{fig:BTO_OTF}c shows the mode evolution during the simulation. In the range that shown in Figs. \ref{fig:BTO_OTF}a-c, about \num{35000} MD steps are taken and only 36 FP calculations are performed.
The fitting process is further taken on $2\times 4\times 4$ supercell (160 atoms) to get the parameters corresponding to Figs. \ref{fig:BTO_OTF}d and e.

Figures \ref{fig:BTO_OTF}d and e show the phase diagrams of \ce{BaTiO3} obtained from the effective Hamiltonian simulation with parameters from the conventional parameterization and from on-the-fly learning, respectively. The supercell size is chosen to be $12\times 12\times 12$ (corresponding to 8640 atoms).
At high temperature, all components of the dipolar mode are zero, characterizing an paraelectric cubic (C) phase. With the decreasing of temperature, the C phase sequentially transforms into ferroelectric tetrahedral (T), orthogonal (O) and rhombohedral (R) phases, characterized by one, two, and three non-zero components of the dipolar mode, respectively. Such C-T, T-O, and O-T phase transition sequence simulated by the effective Hamiltonian with both sets of parameters are correctly reproduced and are consistent with experimental results \cite{BTO_Zhong1995,BTO_PhysRevLett.73.1861}. 
For the calculations with the parameters from conventional FP calculations (Fig. \ref{fig:BTO_OTF}d), the C-T, T-O, and O-T  phase transition temperatures are about 280, 230 and 200 K, respectively. While for the calculation with on-the-fly learning parameters,  they are 380, 270, and 220 K, respectively, much closer to the experimental values of 403, 278 and 183 K, respectively \cite{BTO_PhysRevLett.73.1861}.
Note that such improvement of critical temperature mainly originates from the inclusion of new anharmonic intersite interactions, as discussed in Supplementary Information.

\paragraph*{Solid-solution \ce{Pb(Zr_{0.75}Ti_{0.25})O3}.}
The on-the-fly learning approach is then applied to the solid solution of ferroelectric  \ce{Pb(Zr_{1-x}Ti_{x})O3}, which is of great interest because of its high piezoelectricity and widespread applications \cite{piezo_book}.
We choose the solid solution of \ce{Pb(Zr_{1-x}Ti_{x})O3} ($x$=0.25)  (PZT25) to demonstrate our on-the-fly learning effective Hamiltonian. The active learning is performed on PZT25 using $2\times 2\times 2$ and $2\times 4\times 4$ (40 and 160 atoms, respectively) with random arrangements of Ti and Zr atoms (see Supplementary Information). Using the parameters from this active learning, effective Hamiltonian calculations with supercell of $12\times 12\times 12$ show that PZT25 possesses cubic (C), rhombohedral with space group $R3m$, and rhombohedral with space group $R3c$ phases as the temperature decreases from 700 K to 20 K, with the transition temperature around 540 K (C-$R3m$) and 340 K ($R3m$-$R3c$) (see Fig. \ref{fig:PZT_PT} in the Supplementary Information), very close to the experimental values of 593 K (C to $R3m$) and 390 K (C to $R3c$) \cite{PZT_10.1063/1.123756}, indicating the validity of our on-the-learning scheme for solid-solution structures.

\paragraph*{Prediction and experimental validation of polar skyrmions.}
The emergent and exotic phase of polar topological configurations such as polar skyrmions have garnered enormous interest in condensed-matter physics. Most polar topological configurations were found near the interface of polar superlattices or heterostructures \cite{skyr_Han2022,skyr_Das2019}. 
Here, we find polar skyrmion in \ce{SrTiO3}/\ce{PbTiO3} bilayer by our on-line-fly learning effective Hamiltonian.

The structure of \ce{PbTiO3} with surface capping by  a few \ce{SrTiO3} layers is considered. Supercell of $48 \times 48 \times 48$  with 43 \ce{PbTiO3} layers, 5 \ce{SrTiO3} layers and vacuum layers along the $z$ direction is used. 
The parameters of effective Hamiltonian is obtained by performing our on-the-fly learning approach on \ce{(Pb_{7/8}Sr_{1/8})TiO3} solid solutions with random arrangements of \ce{Pb} and \ce{Sr} atoms. Figure \ref{fig:skyr}a shows the local dipole configuration averaged over the top 10 \ce{PbTiO3} layers  obtained from hybrid Monte Carlo (HMC) simulations at 10 K. One can clearly see that there are  topological upwards-oriented nanodomains embedded in an downwards-oriented matrix. The in-plane polarization within such nanodomains have a center-divergent character with the two-dimensional winding number equal to one of each domain \cite{skyr_Han2022,topo_def_RMP1979}. Figure \ref{fig:skyr}b shows the local dipoles in (010) plane of the nanodomain  delimited by white circle of Fig. \ref{fig:skyr}a, indicating the center-divergent polar N$\acute{e}$el skyrmion (see the sketch in Fig. \ref{fig:skyr}c).


To validate the prediction of the polar skyrmion, the \ce{PbTiO3} film with surface capping 2 nm \ce{SrTiO3} layers is deposited by pulsed laser deposition technique, and the domain configuration is characterized by vector piezoelectric force microscope (PFM), as shown in Fig. \ref{fig:skyr}d-h.
Figure \ref{fig:skyr}d shows a ring-shaped dark contrast in the vertical PFM amplitude image, while a 180° phase contrast with the surrounding background region. Figure \ref{fig:skyr}e shows that there is a clear dark line in the lateral PFM amplitude image and a half-violet and half-yellow contrast in the lateral PFM phase image, indicating a phase inversion of the lateral polarization component along the direction perpendicular to the cantilever. Figures \ref{fig:skyr}f-h show the lateral PFM phase images acquired with the cantilever aligned by 30°, 60°, and 90°, respectively, with respect to the [010] direction. By rotating the sample clockwise for a set of given angles, the dark line and 180° phase inversion of the polarizations in the lateral PFM image rotates continuously with the cantilever, indicating the rotation symmetry about the center of the nanodomain. 
Such observation  confirms the prediction of polar skyrmion in the structure of \ce{PbTiO3} with capping \ce{SrTiO3} on the surface, indicating the validity of our on-the-fly learning effective Hamiltonian approach for such complex interaction and complex system.


\section{Discussion}

\begin{figure}
  \includegraphics[width=\figurewidthhalf]{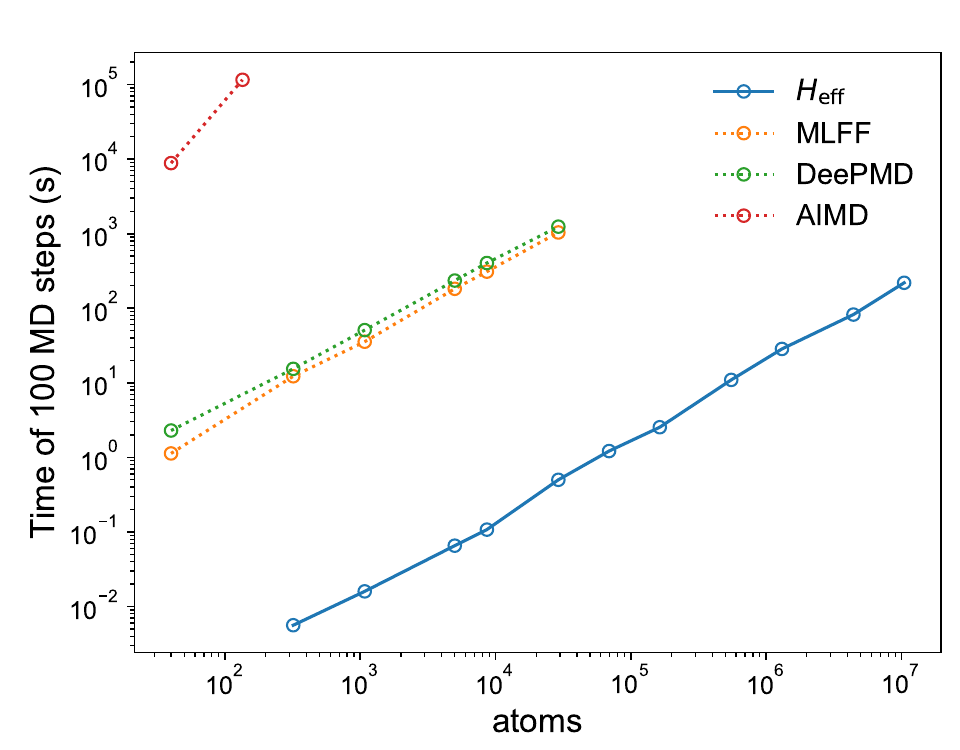}
  \caption{\label{fig:eff_test}
  Computational time for 100 MD steps calculations as a function of the number of atoms in the simulated of \ce{BaTiO3} supercell, using the effective Hamiltonian ($H_{\text{eff}}$),  MLFF, eep potential MD,and ab-initio MD (AIMD).
  The tests are performed on Intel(R) Xeon(R) Silver 4210R CPU using one core, expect for the AIMD simulation, which is performed on the Intel(R) Xeon(R) CPU E5-2680 v3 CPU using 24 cores.
  }
\end{figure}

To demonstrate the computational efficiency of our effective Hamiltonian methods, we compare the time consumed by the effective Hamiltonian MD with other methods, such as deep potential MD \cite{DPMD_PhysRevLett.120.143001}, MLFF MD and ab-initio MD simulations \cite{VaspMLFF_PhysRevB.100.014105}. As shown in Fig. \ref{fig:eff_test}, the time consumed by ab-initio MD simulations increases drastically with the increasing of supercell size, and is much slower than other methods, as consistent with common beliefs. The time spent by other methods increase slowly with the increasing of supercell size within similar slope in the log-log scale. For the same supercell size, the time spent by effective Hamiltonian simulation is less than the deep potential MD and MLFF MD by about 3 orders of magnitude, respectively. Note that these methods possess similar accuracy as they are all based on FP calculations.

In summary, on-the-fly active learning scheme is developed to obtain the parameters of the effective Hamiltonian methods. The parameters are computed during MD simulations. The energy, forces and stress as well as their Bayesian errors are computed at each MD step based on the effective Hamiltonian, and FP calculations are called to fit the parameters when the Bayesian errors are large. The fitting procedure based on Bayesian linear regression provides not only the values of the parameters, but also their uncertainties. 
Such learning scheme offers a new way with high precision to parametrize the effective Hamiltonian in a universal and  automatic process, and is especially highly applicable for the systems that have complex interactions in complex system.


\section{Methods}

\subsection{Effective Hamiltonian for perovskites}
In the effective Hamiltonian of perovskites, the local modes are
(1) the local dipolar mode $\bm u_i$ in each five-atom perovskite unit cell $i$, which is directly proportional to the local electric dipole in unit cell $i$ \cite{BTO_Zhong1995}; (2) the pseudovector $\bm \omega_i$ centered at $B$ site, characterizing the \ce{BX6} octahedral tilting, also known as antiferrodistortive (AFD) distortions \cite{PZT_PhysRevLett.97.157601}; and (3) the local variable $\bm v_i$ characterizing the inhomogeneous strain around the unit cell $i$ \cite{BTO_Zhong1995}.
Note that the $\bm u_i$ and $\bm v_i$ vectors could be chosen to be centered at either $A$ site or $B$ site for different materials.
The potential energy of perovskites contains four main parts
\begin{equation}
  \begin{split}
    E_{\text{pot}}=&
    E_{\text{single}}(\lbrace \bm u_i \rbrace, \lbrace \bm \omega_i \rbrace, \lbrace \bm v_i \rbrace) + \\
    & E_{\text{strain}} (\lbrace \bm u_i \rbrace, \lbrace \bm \omega_i \rbrace, \lbrace \bm v_i \rbrace, \eta) + \\
    & E_{\text{inter}}(\lbrace \bm u_i \rbrace, \lbrace \bm \omega_i \rbrace, \lbrace \bm v_i \rbrace ) + \\
    & E_{\text{spring}} (\lbrace \bm u_i \rbrace, \lbrace \bm \omega_i \rbrace, \lbrace \bm v_i \rbrace, \lbrace\sigma_i\rbrace).
  \end{split}
  \label{eq:Epot}
\end{equation}
The first two terms $E_{\text{single}}$ and $E_{\text{strain}}$ contains mainly the terms already reported in previous effective Hamiltonian works \cite{BTO_Zhong1995,CPI_Chen2020} (with a small amount of extension, see the Supplementary Information for more details).
The last two terms $E_{\text{inter}}$ and $E_{\text{spring}}$  are different from previous works (see, e.g., Refs. \cite{PZT_PhysRevLett.97.157601,BZT_2012PhysRevLett.108.257601,VCA_PhysRevB.61.7877}). Such terms are derived directly from symmetry here, making them more general, and the accuracy of the effective Hamiltonian can then be improved systematically. 

The $E_{\text{inter}}$ in Eq. \eqref{eq:Epot} contains several two-body interaction terms that takes the following form
\begin{equation}
  E_{\text{inter}}^{pq}=\sum_{ijab} p_a (\bm R_i) q_b (\bm R_j) K_{ab} (\bm R_i-\bm R_j),
  \label{eq:Econv}
\end{equation}
where $\bm R_i$ and $\bm R_j$ are the position of the sites indexed by $i$ and $j$, $p$ and $q$ are two variables participating in this interaction, and $a$, $b$ are their subscripts. 
The interaction matrix $K_{ab}(\bm R_i-\bm R_j)$ contains the symmetry and parameter of the interaction. The specific form of the interaction matrix is determined by finding the symmetry invariant terms under the symmetry operations of the reference structure (see Ref. \cite{zhao_energetic_2022}).
This interaction term [Eq. \eqref{eq:Econv}] may be either long-ranged or short-ranged. For long-range interactions, both the $i$ and $j$ indexes run over all the sites in the simulated supercell. On the other hand, for short-range interactions, the $i$ index runs over all the sites in the simulated supercell, while the $j$ runs over the neighbor sites around $i$ (within certain range for each type of interaction). In such case, the interaction matrix is localized.
Note that, the interaction variable $p, q$ here could be not only the primitive degrees of freedom (i.e. $\bm u, \bm v, \bm \omega$), but also their onsite direct products. For example, the 6-dimension vector $\bm U$ expressing the onsite direct product of $\bm u \otimes \bm u$ with subscript $a$ being Voigt notation ($U_1=u_x^2, U_4=u_y u_z$) could be a valid interaction variable in Eq. \eqref{eq:Econv}.
Throughout this article, the expression ``$p^m-q^n$ interaction'' denotes the interaction that includes $m$th order contribution from $p$ and $n$th order contribution from $q$. For example, the ``$u^1-\omega^2$ interaction'' is the interaction that equivalent to
\begin{equation}
  E_{\text{inter}}^{u^1-\omega^2}=\sum_{ij\alpha\beta\gamma} u_\alpha(\bm R_i) \omega_\beta(\bm R_j) \omega_\gamma(\bm R_j) K_{\alpha\beta\gamma} (\bm R_i-\bm R_j).
\end{equation}
As in previous MD and HMC works \cite{HMC_prokhorenko_large_2018}, the interaction in Eq. \eqref{eq:Econv} could be handled in the reciprocal space by using fast Fourier transformation to improve the computational efficiency.

The detailed interaction terms that are used in this work are listed in the Supplementary Information. A brief discussion about the ``inhomogeneous strain'' $\eta_I$ introduced in previous works \cite{BTO_Zhong1995,CPI_Chen2020} is also given in the Supplementary Information.

The $E_{\text{spring}}$ term in Eq. \eqref{eq:Epot} consists of several so-called ``spring'' terms that take the following form
\begin{equation}
  E_{\text{{spring}}}^p=\sum_{ija} p_a(\bm R_i) J_a(\sigma_j, \bm R_j-\bm R_i),
  \label{eq:Eloc_spring}
\end{equation}
where $p$ is a variable [like that in Eq. \eqref{eq:Econv}] that can be primitive local modes ($\bm u, \bm v, \bm \omega$) or their onsite direct products, and $J_a(\sigma_j, \bm R_j-\bm R_i)$ is the interaction matrix containing the symmetry and parameter that depending on the occupation on site $j$ and the position difference between $i$ and $j$ sites. To determine the specific form of $J$ matrix, the $\sigma$ variable is treated as an onsite scalar variable that is invariant under any symmetry operations. Then, the interactions allowed by symmetry are found by performing symmetry operations (of the reference structure space group) on the products $\bm p(\bm R_i) \sigma (\bm R_j)$ and finding the invariant terms under such operations.
Practically, the following spring interactions are considered:
(1) The spring interaction of $\bm u$ of first, second and third order. (2) The spring interaction of $\bm v$ of first order. (3) The spring interaction of $\bm\omega$ with second order. Note that for both the cases of multi A- or B-site element (i.e. the $\sigma$ variables are centered on A- or B-site) and $\omega$ centering on B site, the first order of spring interaction is forbidden by symmetry. Thus, the second order interaction is the lowest order one.
A brief discussion about the relations and differences of the treatment of ``alloying effect'' (using the terminology of Ref. \cite{PZT_PhysRevLett.84.5427}) between previous works and current work is given in the Supplementary Information.

Note that, for a specified materials, some of the above terms  may not be used since their effects are not important.

\subsection{Mode and basis}

The mode is the local collective displacement of atoms in a specified pattern (see Fig. \ref{fig:schematic}a), which is also called lattice Wannier function (LWF) \cite{BTO_Zhong1995,LWF_PhysRevB.52.13236}. In perovskites, the LWF basis of local dipole motion $\bm u$ is typically chosen to be the local phonon mode having $\Gamma_{15}$ symmetry centered on A or B site \cite{1a}. Typically, the LWF basis is determined from the eigenvector associated with the dipolar mode of the force constant or dynamic matrix of cubic perovskite, which takes the form $\bm \xi=(\xi_A, \xi_B, \xi_{X1}, \xi_{X2})$. For example, the displacement of local mode motion $u_{i\alpha}$ centered at $B$ site consisting of the displacement of center $B$ atom by amounts of $u_i\xi_B$, the displacement of eight neighbor $A$ atom by amounts of $u_i\xi_A/8$, and the displacement of the six neighbor $X$ atom by amounts of $u_i\xi_{X1}/2$ or $u_i \xi_{X2}/2$, all along the $\alpha$ direction. Note that it is also possible to get the LWF basis by fitting against the atomic displacement between the reference structure and low energy structure \cite{PTO_PhysRevB.88.064306}.
The local motion $\bm v$ is similar to $\bm u$ but with the basis corresponding to the translation motion of all the atoms in the unit cell, i.e. with $\xi_A=\xi_B=\xi_{X1}=\xi_{X2}$.

The AFD mode $\bm \omega$ is kind of different from the $\bm u$ and $\bm v$ modes, since a neighboring \ce{BX6} octahedron shares the same $X$ atom, and thus the $\bm \omega$ modes are not completely independent from each other. The actual movement of the $X$ atom shared by the $i$ and $j$ sites associated with the AFD mode is given by
\begin{equation}
  \Delta \bm r_X=\frac{a_0}{2}  \hat{\bm {R}}_{ij}\times (\bm\omega_i-\bm\omega_j),  \label{eq:AFD_O_move}
\end{equation}
where $\hat{\bm{R}}_{ij}$ is the unit vector jointing the site $i$ and site $j$. By definition of Eq. \eqref{eq:AFD_O_move}, there are multiple (actually, infinite) different sets of $\lbrace\bm\omega\rbrace$ modes representing the same atomic structure (i.e. with the same set of atomic displacement) \cite{AFDdiff_Ferroelectrics_doi:10.1080/00150199808009158}. For example, it is clear that adding an arbitrary amount of $\bm \omega_0$ to all of the AFD modes does not change the displacement $\Delta \bm r_X$, since the displacement only depends on the \emph{difference} between $\bm \omega_i$ and $\bm \omega_j$.
To eliminate such arbitrariness, we typically impose the following extra restrictions on the AFD vectors and its cyclic permutations:
\begin{equation}
  \forall x_0,~ \sum_{i, n_x(i)=x_0} \omega_{i,x}=0,  \label{eq:AFD_constr}
\end{equation}
where $i$ is the index of unit centered at $n_x(i)\hat x a_0 + n_y(i)\hat y a_0 + n_z(i)\hat z a_0$, $\hat x, \hat y, \hat z$ are unit vectors along the $x,y,z$ axes, and $a_0$ is the lattice constant of the five-atom perovskite unit cell. The summation runs over all the sites in the same layer marked with $n_x(i)=x_0$.
Note that our definition of atomic displacement [Eq. \eqref{eq:AFD_O_move}] is identical to that in Ref. \cite{AFDdiff_Ferroelectrics_doi:10.1080/00150199808009158} [Eq. (1) there in], but our formalism is different from that of Ref. \cite{AFDdiff_Ferroelectrics_doi:10.1080/00150199808009158} by the extra restrictions [Eq. \eqref{eq:AFD_constr}].

It is clear from above that all of the $\bm u, \bm v$ and $\bm \omega$ modes are linked \emph{linearly} with atomic displacement about the reference structure. For a periodic supercell containing $N=L_x\times L_y\times L_z$ five-atom perovskite unit cells, the relation between the modes and atomic displacements could be written as
\begin{equation}
  \mathbf M \mathbf s = \mathbf x,  \label{eq:mode_convention}
\end{equation}
where $\mathbf s$ is a $9N$ column vector containing the modes $\bm u, \bm v, \bm \omega$ in each unit cell, $\mathbf x$ is a $15N$ column vector containing the atomic displacement of each atom in the supercell, and $\mathbf M$ is the matrix containing the information of LWF basis.
The force acting on the \emph{mode} could then be obtained from the chain rule
\begin{equation}
  f_{s,i}=-\frac{\partial E_{\text{pot}}}{\partial s_i}=-\sum_j\frac{\partial E_{\text{pot}}}{\partial x_j} M_{ji}.
\end{equation}
This equation can be written in matrix form as
\begin{equation}
  \mathbf {f_s} = \mathbf M^T \mathbf {f_x},  \label{eq:force_map}
\end{equation}
where $\mathbf{f_s}$ and $\mathbf{f_x}$ gather the forces acting on the modes and atoms, respectively.

Similar to the second-principle lattice dynamics formalism \cite{SPLD_PhysRevB.95.094115}, the actual atomic coordinates in a supercell with homogeneous strain $\eta_H$ and \emph{atomic} displacement is defined as
\begin{equation}
  \bm r_{lk}=(\mathbbm 1+\bm \eta)(\bm R_l + \bm\tau_k) + \bm x_{lk},
\end{equation}
where $\mathbbm 1$ is the $3\times 3$ identity matrix, $\bm\eta$ is the homogeneous strain (in $3\times 3$ matrix format), $\bm R_l$ is lattice vector corresponding to the unit cell $l$, $\bm \tau_k$ is the coordinate of atom inside the unit cell.
Thus, the stress compatible with that calculated from FP should be obtained using the chain rule
\begin{equation}
  \sigma_m=-\frac{\partial' E_{\text{pot}}}{\partial' \eta_m} = -\frac{\partial E_{\text{pot}}}{\partial \eta_m}
  -\sum_{lk\alpha} \frac{\partial E_{\text{pot}}}{\partial x_{lk\alpha}} \frac{\partial x_{lk\alpha}}{\partial \eta_m},
  \label{eq:can_stress}
\end{equation}
as described in Appendix A of  Ref. \cite{SPLD_PhysRevB.95.094115}. Practically in this work, such relation is used \emph{inversely}. The stress $-\partial' E_{\text{pot}}/\partial' \eta_m$ obtained from the FP calculations are converted to $-\partial E_{\text{pot}}/\partial \eta_m$, compatible with the direct definition of the effective Hamiltonian.

\subsection{Bayesian linear regression}

Given two necessary assumptions satisfied (see Appendix B of Ref. \cite{VaspMLFF_PhysRevB.100.014105}), the posterior distribution of the parameter is a multidimensional Gaussian distribution
\begin{equation}
  p(\mathbf w|\mathbf Y)=\mathcal{N} (\mathbf{\bar w}, \bm \Sigma),
\end{equation}
where the center of the distribution
\begin{equation}
  \mathbf {\bar w}=\frac{1}{\sigma_v^2} \bm \Sigma \bm\Phi^T \mathbf Y  \label{eq:ave_param}
\end{equation}
is the desired optimal parameters, and the variance
\begin{equation}
  \bm\Sigma=\left[ \frac{1}{\sigma_w^2}\mathbf I + \frac{1}{\sigma_v^2} \bm\Phi^T\bm\Phi \right]^{-1}
  \label{eq:param_uncert}
\end{equation}
is a measure of the uncertainty of the parameters. Here, $\mathbf I$ is the identity matrix, $\sigma_v$ is a hyperparameter describing the deviation of FP data from the model prediction $\bm \phi_\alpha \mathbf w$, and $\sigma_w$ is a hyperparameter describing the covariance of the prior distribution of parameter vector $\mathbf w$ (see Appendix B of Ref. \cite{VaspMLFF_PhysRevB.100.014105} for more details).

Given the observation of the training set, the posterior distribution of the energy, forces and stress of a new structure is also shown to be a Gaussian distribution \cite{VaspMLFF_PhysRevB.100.014105}
\begin{equation}
  p(\tilde{\mathbf y}|\mathbf Y)=\mathcal{N}(\bm\phi\mathbf{\bar w}, \bm\sigma),
\end{equation}
where the covariance matrix
\begin{equation}
  \bm\sigma=\sigma_v^2\mathbf I + \bm \phi \bm\Sigma \bm\phi^T  \label{eq:uncert_label}
\end{equation}
measures the uncertainty of the prediction on the new structure. Following Ref.\cite{VaspMLFF_PhysRevB.100.014105}, the diagonal elements of the second term is used as the Bayesian error. If the Bayesian error is large, the prediction on the energy, forces and stress by current effective Hamiltonian model is considered unreliable, then new FP calculation is required to fit the parameters.

The hyperparameters $\sigma_v$ and $\sigma_w$ are determined by evidence approximation \cite{VaspMLFF_PhysRevB.100.014105}, in which the marginal likelihood function corresponding to the probability of observing the FP data $\mathbf Y$ with $\sigma_v$ and $\sigma_w$ is maximized [see Eq. (31) and Appendix C in Ref. \cite{VaspMLFF_PhysRevB.100.014105}]. Practically, the hyperparameters $\sigma_v$ and $\sigma_w$ are calculated along with $\bm \Sigma$ and $\mathbf{\bar w}$ by executing self-consistent iterations at each time when FP data for a new structure is collected.

The Bayesian linear regression described above is equivalent to the ridge regression \cite{MLFF_Review_jinnouchi_-fly_2020,ML_watermelon_zhou} in which the target function
\begin{equation}
  \mathcal{O}=\| \bm\Phi \mathbf w - \mathbf Y \| + \lambda \| \mathbf w \|^2  \label{eq:ridge}
\end{equation}
is minimized, where $\lambda$ is the Tikhonov parameter which is equivalent to $\sigma_v^2/\sigma_w^2$ here \cite{MLFF_Review_jinnouchi_-fly_2020,ML_watermelon_zhou}. The main purpose of imposing the Tikhonov parameter is to prevent overfitting \cite{ML_watermelon_zhou}. However, in the context of effective Hamiltonian parametrization, there are only a small amount of parameters to be determined (typically from several tens to over a hundred), while the number of values collected from FP calculations are typically much larger, which means the linear equations $\bm\Phi\mathbf w=\mathbf Y$ is greatly overdetermined. In such case, the regularization is usually not necessary.
If the regularization term $\lambda \|\mathbf w\|$ in Eq. \eqref{eq:ridge} is removed, the problem becomes a simple linear least square fitting, and the parameter could be simply solved by
\begin{equation}
  \mathbf w=\bm\Phi^+ \mathbf Y,
\end{equation}
where $\Phi^+$ is the Moore-Penrose pseudoinverse of the matrix $\bm\Phi$, which could be computed by performing the singular-value decomposition of $\bm \Phi$ \cite{VaspMLFF_verdi_thermal_2021}.
Indeed, our numerical tests show that the resulting parameter from such fitting without regularization is pretty close to that obtained by Bayesian linear regression. Similar observation is also reported in the context of MLFF with Gaussian approximation potential model (or its analogs) \cite{VaspMLFF_verdi_thermal_2021}.
On the other hand, in effective Hamiltonian, unlike Gaussian approximation potential, the parameters in $\mathbf w$ have different dimensions, and it is hard to balance the \emph{values} between different parameters, indicating that the regularization may be not suitable for effective Hamiltonian.
Based on the two reasons above, in our fitting scheme, it is typically assumed that $\sigma_w\rightarrow \infty$, and thus the equivalent Tikhonov regularization parameter $\lambda$ approaches zero, and the fitting scheme is equivalent to the linear least square fitting.


\subsection{Computational details}
\label{sec:comp_details}

On-the-fly machine learning for parametrization of effective Hamiltonian is performed on $2\times 2\times 2$ or $2\times 4\times 4$ supercells (corresponding to 40 or 160 atoms). The MD simulations are performed with isothermal-isobaric ($NPT$) ensemble using Evans-Hoover thermostat \cite{HeffMD_PhysRevB.77.012102}. Typically, each MD simulation on a given structure is executed for \SI{20}{ps} to \SI{200}{ps}.
For each MD step the FP calculation is required by the on-the-fly machine learning process, first-principles self-consistent calculation within density functional theory (DFT) is performed. All the FP calculations are performed using the VASP package \cite{VASP_PhysRevB.50.17953} with projector augmented wave (PAW) method. The solid-revised Perdew-Burke-Ernzerhof (PBEsol) \cite{PBEsol_PhysRevLett.100.136406} functional is used.
The $3\times 3\times 3$ and $3\times 2\times 2$ k-point meshes are used for the supercells with 40 and 160 atoms, respectively, and the plane wave cutoff of \SI{550}{eV} is employed.
The optical dielectric constant is computed using the density functional perturbation theory (DFPT) \cite{DFPT_PhysRevB.55.10355}. The Born effective charge of the local mode is obtained by fitting the polarization against the local mode amplitude, where the polarization is computed using the Berry phase method \cite{BerryPhase_PhysRevB.47.1651}.

The phase transition simulations are conducted by Monte Carlo (MC) simulations with Metropolis algorithm \cite{Metropolis1953} or hybrid MC algorithm \cite{HMC_prokhorenko_large_2018} (HMC)  with the effective Hamiltonian method. Each HMC sweep consists of 40 MD steps. Supercells of $12\times 12\times 12$ (corresponding to 8640 atoms) are used, unless specially noted. For the phase transition simulations, the systems are cooled down from high temperature  (\SI{450}{K} and \SI{700}{K} for \ce{BaTiO3} and PZT25, respectively) to \SI{20}{K} with relatively small temperature steps of \SI{10}{K}.

In the MD and MC simulations, the following quantities are computed: (i) the average dipolar mode defined as $\bm u=\frac{1}{N}\sum_i \bm u_i$, (ii) the average amplitude of dipolar mode defined as $|u|=\frac{1}{N}\sum_i |\bm u_i|$, and (iii) the AFD at $R$ point defined as
$\bm\omega_R=\frac{1}{N}\sum_i \bm\omega_i (-1)^{n_x(i)+n_y(i)+n_z(i)}$.

In the study of \ce{BaTiO3}, the dipolar mode $\lbrace\bm u_i\rbrace$ is chosen to be centered at B site, the inhomogeneous strain variable $\lbrace\bm v_i\rbrace$ is chosen to be centered at A site, and the AFD $\lbrace\bm\omega_i\rbrace$ variables are frozen at zero since they are not important in \ce{BaTiO3}. Such configuration of the degrees of freedom is consistent with previous works \cite{BTO_Zhong1995}.
In the effective Hamiltonian, the dipolar mode onsite energy [Eq. \eqref{eq:Eself}] is considered up to the quartic order, the elastic energy [Eq. \eqref{eq:EelasH}] is considered up to the quadratic order, and the $\eta-u$ interaction is considered only up to the first term in Eq. \eqref{eq:Eintu}. The $2\times 2\times 2$ and $2\times 4\times 4$ supercells (corresponding to 40 and 160 atoms, respectively) are used in turn for on-the-fly learning.
The parameters associated with Fig. \ref{fig:BTO_OTF}d is obtained using conventional method \cite{BTO_Zhong1995,1a} with PBEsol functional. The $j_5$ and $j_7$ parameters (see Ref. \cite{BTO_Zhong1995}) is set to zero, as in Ref. \cite{feram_nishimatsu_2010_134106}.

In the study of PZT25, the local dipolar mode $\lbrace\bm u_i\rbrace$ is chosen to be centered at A site, the inhomogeneous variable $\lbrace \bm v_i\rbrace$ and the AFD pseudovector $\lbrace\bm\omega_i\rbrace$ are centered at B site. The variable $\lbrace\sigma_i\rbrace$ is introduced to describe the occupation of Zr and Ti atoms at B site, where $\sigma_i=1, 2$ denote Ti, Zr atom sit at the B site indexed by $i$, respectively.
The dipolar mode onsite energy is expanded up to the fourth order. The spring interaction of $u$ is considered up to the third order and the nearest neighbor, the spring interaction of $v$ is considered up to the first order and second nearest neighbor, and the spring interaction of $\omega$ is considered up to the second order (which is the symmetry-allowed interaction of the lowest order) and the nearest neighbor. The Zr and Ti atoms are distributed randomly in the simulated supercells.

In the study of \ce{SrTiO3}/\ce{PbTiO3} bilayer, the effective Hamiltonian is fitted to \ce{(Pb_{7/8}Sr_{1/8})TiO3} solid solutions. The local dipolar mode $\lbrace\bm u_i\rbrace$ is chosen to be centered at A site, the inhomogeneous variable $\lbrace \bm v_i\rbrace$ and the AFD pseudovector $\lbrace\bm\omega_i\rbrace$ are centered at B site. The variable $\lbrace\sigma_i\rbrace$ is introduced to describe the occupation of Pb and Sr atoms at A site, where $\sigma_i=1, 2$ denote Pb, Sr atom sit at the A site indexed by $i$, respectively.
The LWF basis of the local dipolar mode is obtained from fitting against the atomic distortion between ferroelectric tetrahedral phase and cubic perovskite phase. 
The dipolar mode onsite energy is expanded up to the fourth order. The spring interaction of $u$ is considered up to the third order and the nearest neighbor, the spring interaction of $v$ is considered up to the first order and first nearest neighbor, and the spring interaction of $\omega$ is considered up to the second order (which is the symmetry-allowed interaction of the lowest order) and the nearest neighbor. The Pb and Sr atoms are distributed randomly in the supercell during the fitting process.
The \ce{SrTiO3}/\ce{PbTiO3} bilayer is modeled by $48\times 48\times 48$ supercell (corresponding to approximately 552960 atoms) that consists of 43 unit cell layers of \ce{PbTiO3} and 5 unit cell layers of \ce{SrTiO3} along the $z$ axis, where periodic boundary condition is induced in $x, y$ axes but not $z$ axis. Both $(001)$ bilayers are terminated with A site layer. An epitaxy strain of \num{-0.58}\% is imposed to mimic the \ce{SrTiO3} substrate. 
The local configuration of Fig. \ref{fig:skyr}a,b is obtained from a quench simulation (fast cool from 410 K to 10 K with temperature step of 100 K and 5000 HMC sweeps at each temperature).

In the computational efficiency tests for Fig. \ref{fig:eff_test}, the effective Hamiltonian simulation is conducted with one CPU core on different supercell sizes of \ce{BaTiO3} for \num{10000} steps, the average time spent by each MD step is then calculated (with the initial preparation time excluded). For each size of supercell, such process is repeated for 5 times to get the average time.
The MLFF simulation is performed using the VASP \cite{VaspMLFF_PhysRevB.100.014105} package version 6.4.2. The force field for \ce{BaTiO3} is first trained within $2\times 2\times 2$ supercell (40 atoms) at 300 K using 10000 MD steps. In this process, 461 local reference structures are collected. Then, it is switched to the prediction-only mode (ML\_MODE=run) after refitting the field (with ML\_MODE=refit) to measure the consumed time. For each supercell size, the simulation is performed with 1 CPU core for 100 MD steps, and the consumed times by each step (excluding the first and last step) are averaged to produce the results.
The deep potential MD \cite{DPMD_PhysRevLett.120.143001} simulation is performed with the LAMMPS package with one CPU core. Five repeat simulations, each last for 1000 steps, are conducted for each supercell size, and the time spent by each MD step is averaged.
All the tests above are performed on Intel(R) Xeon(R) Silver 4210R CPU using one core.
The {\it ab-initio} MD simulation is performed using the VASP package on $2\times 2\times 2$ and $3\times 3\times 3$ supercells (40 and 135 atoms, respectively) with Gamma-centered K point mesh of $3\times 3\times 3$ and $2\times 2\times 2$, respectively. For each supercell size, the simulation lasts for 100 steps. The time consumed by each step (apart from the first step) is averaged. The {\it ab-initio} MD simulations are performed on the Intel(R) Xeon(R) CPU E5-2680 v3 CPU using 24 cores.

\subsection{Sample deposition}
The \ce{SrTiO3}/\ce{PbTiO3} bilayer heterostructures are deposited by pulsed laser deposition. The \ce{PbTiO3} films, about 40 nm in thickness, were deposited on (001)-oriented \ce{SrTiO3} substrates with 80 nm thick \ce{SrRuO3} electrodes, followed by depositing a 2 nm thick \ce{SrTiO3} capping layer. The \ce{SrRuO3} electrode, \ce{PbTiO3} film and \ce{SrTiO3} capping layer were deposited at 660, 620, and \SI{700}{\degreeCelsius}, respectively, using a 248-nm KrF excimer laser (COMPex Pro 205F, Coherent) with an energy flux density of \SI{1.5}{J/cm^2} on \ce{SrRuO3}, \ce{PbTiO3} and \ce{SrTiO3} ceramic targets and a repetition rate of 3 Hz. 20\% excessive Pb was added into the \ce{PbTiO3} target to compensate the Pb loss during deposition. The oxygen partial pressure for the deposition of \ce{SrRuO3} and \ce{SrTiO3} is 100 mTorr and for the deposition of \ce{PbTiO3} is 80 mTorr.

\subsection{PFM measurement}
Ferroelectric domain structures of various \ce{SrTiO3}/\ce{PbTiO3} bilayers were characterized at room temperature by atomic force microscope (Cypher ES, Asylum Research). NanoWorld EFM platinum/iridium-coated tips and Adama Supersharp Au tips, both 2.8 N/m in force constant, were used in PFM measurements. The ac signal applied on the tip for all the PFM measurements is 800 mV in amplitude. The samples were grounded in all the measurements. Piezoresponse phase-voltage hysteresis loops were collected in the dual a.c. resonance tracking mode. The vector PFM were conducted with different in-plane sample rotation angles to reconstruct the domain structures \cite{PFM_Kim2018,PFM_Kim2019}. 


\begin{acknowledgments}
  We thank the useful discussions with Dr. F. Karsai in VASP Software GmbH. 
  X.M., H.C., Z.Y., Y.Y. and D.W. thank the National Key R$\&$D Programs of China (grant NOs. 2022YFB3807601, 2020YFA0711504), the National Natural Science Foundation of China (grant NOs. 12274201, 51725203, 51721001, 52003117 and U1932115) and the Natural Science Foundation of Jiangsu Province (grant NO. BK20200262). 
  L.B. thanks the Office of Naval Research Grant No. N00014-21-1-2086 and the Vannevar Bush Faculty Fellowship (VBFF) Grant No. N00014-20-1-2834 from the Department of Defense.
  J.I.G. thanks the financial support of the Luxembourg National Research Fund (FNR) through project C21/MS/15799044/FERRODYNAMICS.
  We are grateful to the High Performance Computing Center (HPCC) resources of Nanjing University for the calculations.
\end{acknowledgments}


\bibliography{HeffML}

\end{document}